\documentclass[aps,prc,twocolumn,superscriptaddress,showpacs,nofootinbib,floatfix]{revtex4}
\usepackage{graphicx}
\usepackage{epsfig}
\usepackage{dcolumn}
\usepackage{bm}
\usepackage{amssymb}

\begin{document}
\title{Quark number scaling of $v_2$ in transverse kinetic energy and it's implications for coalescence models}
\newcommand{\columbia}{Columbia University, New York, NY 10027 and Nevis Laboratories, Irvington, NY 10533, USA}
\newcommand{\sunysb}{Department of Chemistry,
Stony Brook University, Stony Brook, NY 11794, USA}
\affiliation{\columbia}
\newcommand{\ornl}{Oak Ridge National Laboratory, Oak Ridge, TN 37831, USA}
\affiliation{\ornl}
\author{Jiangyong Jia}\affiliation{\columbia}\affiliation{\sunysb}
\author{Chun Zhang}\affiliation{\ornl}
\date{\today}
\begin{abstract}
We find that a simple extension of the coalescence model is
sufficient to incorporate the perfect quark number scaling
behavior of the elliptic flow in transverse kinetic energy,
recently discovered by the PHENIX Collaboration. The flavor
dependence of the elliptic flow can be consistently described in
the low and intermediate $p_T$ if the transverse kinetic energy is
conserved in the $2\rightarrow1$ or $3\rightarrow1$ parton
coalescence process at the hadronization. Thus suggesting the
quark coalescence as a possible hadronization mechanism at low
$p_T$ as well.
\end{abstract}

\maketitle

Quark recombination or coalesce
models~\cite{Voloshin:2002wa,Fries:2003kq,Greco:2003mm,Hwa:2002tu,Molnar:2003ff}
have been proposed to explain the large baryon to meson
ratio~\cite{Adler:2003kg,Adams:2003am} and the number of
constituent quark scaling (NCQ-scaling) of the hadron $v_2$ in
$p_T$~\cite{Adler:2003kt,Adams:2003am} seen in RHIC data. Since
coalescence models explicitly assume partonic degree of freedom,
their apparent, albeit qualitative, success in describing the data
at intermediate $p_T$ is a strong support of the idea that the
partonic matter has been created in Au-Au collisions at RHIC.

Coalescence models calculate the meson production via the
convolution of Wigner functions for constituent quarks and the
meson~\cite{Fries:2003kq,Greco:2003mm,Hwa:2002tu,Molnar:2003ff}:

\begin{equation}
\label{eq:1} \frac{d^3N}{dp^3} = \int \prod_{i=1}^2
d^3x_id^3p_iF_M(x_1,p_1,x_2,p_2)W_{M}(p,p_1,p_2,x_1,x_2)
\end{equation}
where F is the joint $q$ and $\bar{q}$ phase space distribution,
$W_{M}$ is the meson Wigner function, which is typically
approximated by
\begin{equation}
\label{eq:2} W_{M} =
\Phi_{M}(x_1-x_2,p_1-p_2)\delta(p_T-p_{T,1}-p_{T,2})
\end{equation}
The $\delta$ function enforce the momentum conservation. It is
straightforward to generalize these formula for baryons.

There are several implicit assumptions in coalescence models. Only
valence quark degrees of freedom are assumed in the coalescence
calculation, and the dynamic gluon contribution (appears as higher
Fock state)~\cite{Muller:2005pv} could lead to a 20\% correction
to the NCQ-scaling. There is a unitarity problem at low
$p_T$~\cite{Fries:2003kq,Molnar:2003ff} due the quadratical
dependence of the meson yield on the number of quarks, although
this can be reconciled when one take into a finite freeze out time
and no quarks are used in more than one hadron~\cite{Yang:2005dg}.
In addition, coalescence process reduce the entropy. This can be
partially resolved by including resonance decay of unstable
hadrons~\cite{Greco:2004ex}.

As Eq.\ref{eq:1} indicates, the level of NCQ-scaling depends on
both the final state parton phase space distribution, described by
$F_{M}$, and the hadronization process which is encoded in
$W_{M}$. The constituent quark phase space distributions are
driven by the space-time evolution of the partonic matter, which
is typically described by hydrodynamics or parton cascade.
Coalescence models explicitly assume that $F_{M}$ can be
factorized into the single quark distribution function,
$F_{M}=f(x_1,p_1)f(x_2,p_2)$, with identical anisotropic phase
space distribution (only $v_2$ is considered here.)
\begin{equation}
f(x,p) = n(x,p)\left(1+2v_{2,q}(x,p)\rm{cos}(2\phi)\right).
\end{equation}
For a naive coalescence model where $q$ and $\bar{q}$ are assumed
to be co-moving : $\Phi_{M} \propto
\delta^3(x_1-x_2)\delta^3(p_1-p_2)$, the spatially averaged $v_2$
for quark and meson is~\cite{Molnar:2005wf}:
\begin{eqnarray}
\overline{v_{2,q}}  &=& \frac{\int d^3xd\phi
f(x,\frac{p}{2})\rm{cos}(2\phi)}{\int d^3x d\phi
f(x,\frac{p}{2})}\\\nonumber&=&
\frac{\left<n(x,\frac{p}{2})v_{2,q}(x,\frac{p}{2})\right>_x}{\left<n(x,\frac{p}{2})\right>_x}
\\
\overline{v_{2,M}} &=& \frac{\int d^3xd\phi
f^2(x,\frac{p}{2})\rm{cos}(2\phi)}{\int d^3x d\phi
f^2(x,\frac{p}{2})}\\\nonumber&=&
\frac{2\left<n^2(x,\frac{p}{2})v_{2,q}(x,\frac{p}{2})\right>_x}{\left<n^2(x,\frac{p}{2})\left(1+2v^2_{2,q}(x,\frac{p}{2}\right)\right>_x}
\end{eqnarray}
where $\left<A(x)\right>_x = \int d^3xA(x)$. Because quark density
$n(x,p)$ and parton anisotropy $v_{2,q}(x,p)$ could depend on
quark local coordinate and momentum, NCQ-scaling in general does
not follow as pointed in~\cite{Pratt:2004zq,Molnar:2005wf}. The
space momentum correlation of the parton system before
hadronization requires detailed study of the dynamical evolution
of the system~\cite{Molnar:2004rr,Molnar:2005wf}.

The level of NCQ-scaling also depends on the modelling of
hadronization process. It is known that the finite width of the
hadron internal wave function can lead to 10-20\% violation of the
$v_2$ NCQ-scaling~\cite{Greco:2005jk}. In the extreme case where
wave function is very wide, NCQ-scaling can be destroyed
completely~\cite{Lin:2002rw}. Since coalescence deals with
$2,3\rightarrow1$ processes, in general quarks or the coalescenced
hadron need to be off mass shell~\cite{Scheibl:1998tk}, and the
interactions with the surrounding medium are necessary to
neutralize their virtuality. Thus the constituent quarks should be
treated as quasi-particles and there is no reason a priori to
assume the momentum conservation in coalescence process
(Eq.\ref{eq:2}). Our goal is to study the consequences of
modifying this assumption.

\begin{figure}
\epsfig{file=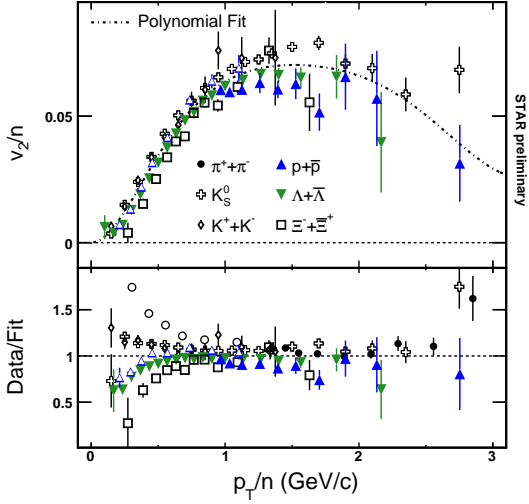,width=0.8\columnwidth}
\caption{\label{fig:1} (Color online) Figure taken from
~\cite{Oldenburg:2005er} a) A compilation of $v_2/n$ as function
of $p_T/n$. b) The ratio to a common polynomial function.}
\end{figure}

If coalescence is a relevant hadronization mechanism, it should be
valid at all momentum. However, from the data point of view,
NCQ-scaling of the $v_2$, when plotted in $p_T$, is not
perfect~\cite{Oldenburg:2005er}, i.e. $v_2(p_T/n)/n$ is not a
universal curve. This is illustrated by Fig.\ref{fig:1}, where the
ratios of the scaled elliptic flow ($v_2/n$) for various hadrons
to a combined polynomial fit are plotted as function of scaled
transverse momentum ($p_T/n$). NCQ-scaling only works at 20\%
level at $p_T>2$ GeV/$c$, then it breaks badly at $p_T <2$
GeV/$c$. The breaking has a systematic dependence on the mass. It
undershoots the $v_2$ values for light mesons such as pion and
overshoots the $v_2$ for heavy baryons such as $\Xi$. Attempts
have been made to reconcile the breaking of pion $v_2$ by
including the resonance effect in the coalescence
models~\cite{Dong:2004ve,Greco:2004ex}. The agreement with data
can be achieved qualitatively. We are not aware of any mechanisms
within existing coalescence models that can accommodate the
deviation of heavy baryons.
\begin{figure}
\epsfig{file=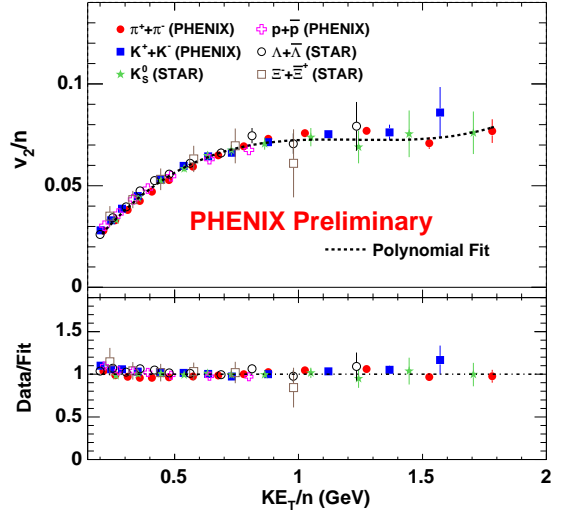,width=0.9\columnwidth}
\caption{\label{fig:2} (Color online) Figure taken
from~\cite{phenix}: a) A compilation of $v_2/n$ as function of
$\rm{KE}_T/n$. b) The ratio to a common polynomial function.}
\end{figure}

Recently it was point out by PHENIX Collaboration~\cite{phenix}
that the constituent quark scaling of $v_2$ can be preserved at
low $p_T$ if hadron transverse kinetic energy $\rm{KE}_T = m_T-m$
instead of $p_T$ is used as a scaling variable. We duplicated the
PHENIX finding in Fig.\ref{fig:2}. The deviation from the
universal $v_2(\rm{KE}_T/n)/n$ scaling curve is very small, less
than 10\% in all cases. Motivated by this observation, we modify
the momentum conservation relation into kinetic energy
conservation relation in the coalescence formula Eq.\ref{eq:2}:
\begin{equation}
\label{eq:6}
 W_{M} = \Phi_{M}(x_1-x_2,p_1-p_2)\delta(m_T-m_{T,1}-m_{T,2} -m + m_1 + m_2)
\end{equation}
Without very detailed calculations, we can make the following
remarks.

\begin{figure}
\epsfig{file=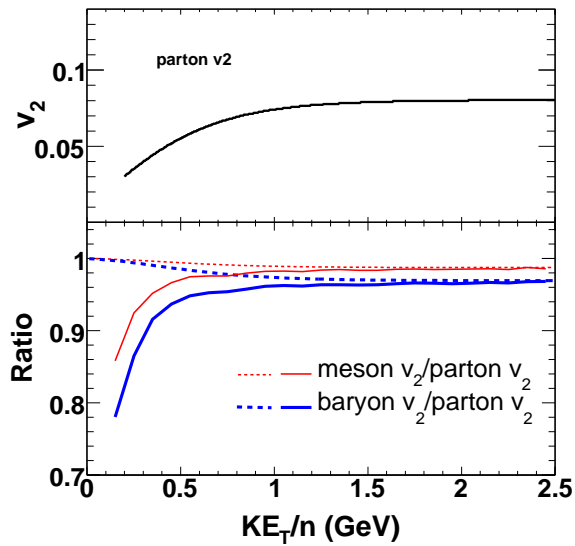,width=0.9\columnwidth}
\caption{\label{fig:3} (Color online) a) The input quark
$v_2(\rm{KE}_T)$. b) The ratio of the scaled $v_2$
($v_2(\rm{KE}_T/n)/n$) for meson and baryon over quark $v_2$
(solid lines) and ratios of the scaled $v_2$ from Eq.\ref{eq:7}
and Eq.\ref{eq:8} over quark $v_2$ (dashed lines).}
\end{figure}

$\it{1.}$ For the naive coalescence scenario, where only the
co-moving $q$ and $\bar{q}$ recombine into hadron, assuming
spatially uniform quark density $n(x,p) = n(p)$ or flow anisotropy
$v_{2,q}(x,p) = v_{2,q}(p)$, we obtain the usual NCQ-scaling
relation for meson and baryon~\cite{Fries:2003kq,Molnar:2003ff},
except it is in $\rm{KE}_T$, instead of $p_T$.

\begin{eqnarray}
\label{eq:7} &&v_{2,M}(\rm{KE}_T) =
\left. {\frac{2v_{2,q}}{1+2v^2_{2,q}}}\right|_{\frac{\rm{KE}_T}{2}} \approx 2v_{2,q}(\frac{\rm{KE}_T}{2})\\
\label{eq:8} &&v_{2,B}(\rm{KE}_T) =
\left.{\frac{3v_{2,q}+3v^3_{2,q}}{1+6v^2_{2,q}}}\right|_{\frac{\rm{KE}_T}{3}}\approx
3v_{2,q}(\frac{\rm{KE}_T}{3})
\end{eqnarray}
We have confirmed these relations through a simple Monte-Carlo
simulation of the naive coalescence process, where $q$ or
$\bar{q}$ are assumed to have mass of 350 MeV and an exponential
spectra in $m_T$. In addition, similar to what is done in other
recombination models, the transverse momentum of the combining
quarks in x or y direction are required to be within 0.24 GeV/$c$
of each other.
In essence the coalescence formula Eq.\ref{eq:1} becomes
\begin{eqnarray} \label{eq:9} \nonumber\frac{d^3N}{dp^3}
&=& \int
d^2p_1d^2p_2e^{-(m_{T,1}+m_{T,2})/A}\delta(\rm{KE}_T-\rm{KE}_{T,1}-\rm{KE}_{T,2})
\\&&\Theta(\Delta_p-|p_{x,1}-p_{x,2}|)\Theta(\Delta_p-|p_{y,1}-p_{y,2}|)
\end{eqnarray}
where A = 0.17 GeV and $\Delta_p=0.24$ GeV/$c$ is the cut off of
relative momentum between the two quarks. Since we focus on
mid-rapidity, longitudinal momentum direction has been ignored in
this study. Results of our calculations are shown in
Fig.\ref{fig:3}. The calculations are also compared with the
analytical results given by Eq.\ref{eq:7} and \ref{eq:8} in dashed
lines. Clearly, the breaking of scaling due to the finite momentum
spread in coalescence become significant only at KE$_T<1$ GeV.
Above 1 GeV, both curves approach the value indicated by
Eq.\ref{eq:7} and \ref{eq:8} (indicated by dashed lines). However,
the split between meson and baryon in $\rm{KE}_T/n$ would not be
discernable given the precision of the data shown by
Fig.\ref{fig:2}. We noted that Eq.\ref{eq:7} and Eq.\ref{eq:8} are
valid as long as the constituent quark spectra are exponential in
$m_T$, independent of the quark mass.

$\it{2.}$ The kinetic energy conservation condition is more than a
simple improvement over the previous coalescence models only at
low $p_T$. It represents a quantitatively different recombination
condition at all $p_T$. At large $p_T$ ($p_T\gg m$), coalescence
condition Eq.\ref{eq:6} becomes $p_T-p_{T,1}-p_{T,2} = m
-m_1-m_2$. It is different from momentum conservation, because in
general $m \neq m_1+m_2$. This can be clearly seen by comparing
the bottom panels of Fig.\ref{fig:1} and Fig.\ref{fig:2}. Going
from $p_T/n$ to $\rm{KE}_T/n$, one effectively shift high $p_T$
part of light meson $v_2/n$ to the right and the low high $p_T$
part of heavy baryon $v_2/n$ to the left by an amount on the order
of the hadron mass. However, the resulting scaled $v_2$ curves
falls on top of each other.

$\it{3.}$ In hydrodynamic model with Cooper-Fryer freeze
out~\cite{Kolb:2000sd}, hadrons are statistically produced in the
local co-moving frame with certain radial flow. At low $p_T$, the
elliptic flow should be proportional to $p_T^2$ according to the a
general analytical expansion~\cite{Danielewicz:1994nb}. The
Buda-Lund hydro model~\cite{Csanad:2003qa} suggests a universal
elliptic flow scaling : $v_2 = I_1(w)/I_0(w), w\propto
p_T^2/2m_T$, which, at low $p_T$ (thus small $w$), reduce to $v_2
\approx 0.5w \propto p_T^2/2m=\rm{KE}_T$. We also arrive this
relation through a schematic derivation~\cite{jjia:derivation}.

For any variable $X$ and a linear relation between elliptic flow
and $X$ : $v_2 \propto X$, NCQ-scaling in $X$ is a trivial
additive artifact: $v_2/n \propto X/n$. It was realized
by~\cite{Pratt:2004zq,Molnar:2004rr} that either hydrodynamical
mass ordering or NCQ-scaling of $v_2$ can results from quark
coalescence. In our point of view, the coalescence condition
Eq.\ref{eq:6} provide a natural reconciliation between the
statistical hadronization and hadronization via coalescence. Since
$v_2$ is proportional to $\rm{KE}_T$ at low $p_T$, $v_2$ for
various hadrons naturally fall on a universal curve in small
$\rm{KE}_T$ independent of whether they are scaled by the number
of constituent quark or not.

Due to the non-linear relation between $p_T$ and $\rm{KE}_T$, at
low or intermediate $p_T$ region where mass effect is important,
NCQ-scaling can not hold in both $p_T$ and $\rm{KE}_T$
simultaneously. At $p_T\gg m$, a schematic hydro calculation gives
$v_2\approx \tanh(\xi p_T)$~\cite{Huovinen:2001cy}, where $\xi$
depends on flow velocity. However, this relation gives neither
NCQ-scaling in $p_T$ nor NCQ-scaling in $\rm{KE}_T$.

A complete model should be able to describe not only elliptic flow
but also the particle ratio, single particle spectra, etc. Our
study is rather limited in this regard. We do not attempt to
describe any observables other than elliptic flow. Various
hadronic effects after quark coalescence, such as the hadronic
re-scattering and resonance decay, are ignored in current study.
Our discussion is focused on the physical consequences when quark
coalescence is extended to low $p_T$. We have demonstrated that
results from a simple coalescence approach agree with the data
fairly well in KE$_T$. The deviation of the pion $v_2/n$ from
$p_T/n$ scaling is naturally resolved when plotted in KE$_T$/n
variable, without the need to rely on the resonance effect. The
fact that RHIC data indicate strong flavor dependence of the $v_2$
and strong violation of the NCQ-scaling of the $v_2$ in $p_T$
suggests that $\rm{KE}_T$ scaling of the $v_2$ at intermediate
$p_T$ is a rather non-trivial observation. It suggests that the
quark coalescence might be an important hadronization mechanism at
low $p_T$ ($p_T<2$ GeV/$c$).

We thank M.~Issah for pointing us to the Fig.2 in his
talk~\cite{phenix}. Discussions with R.~Fries, C.~M.~Ko are
appreciated.


\begin{references}
\bibitem{Voloshin:2002wa}
  S.~A.~Voloshin,
  Nucl.\ Phys.\ A {\bf 715}, 379 (2003)

\bibitem{Fries:2003kq}
  R.~J.~Fries, B.~Muller, C.~Nonaka and S.~A.~Bass,
 ``Hadronization in heavy ion collisions: Recombination and fragmentation  of
  Phys.\ Rev.\ Lett.\  {\bf 90}, 202303 (2003),
  Phys.\ Rev.\ C {\bf 68}, 044902 (2003)
\bibitem{Greco:2003mm}
  V.~Greco, C.~M.~Ko and P.~Levai,
  Phys.\ Rev.\ Lett.\  {\bf 90}, 202302 (2003),
  Phys.\ Rev.\ C {\bf 68}, 034904 (2003)
\bibitem{Hwa:2002tu}
  R.~C.~Hwa and C.~B.~Yang,
  Phys.\ Rev.\ C {\bf 67}, 034902 (2003)
\bibitem{Molnar:2003ff}
  D.~Molnar and S.~A.~Voloshin,
  Phys.\ Rev.\ Lett.\  {\bf 91}, 092301 (2003)
\bibitem{Adler:2003kg}
  S.~S.~Adler {\it et al.}  [PHENIX Collaboration],
  Phys.\ Rev.\ Lett.\  {\bf 91}, 172301 (2003)
\bibitem{Adams:2003am}
  J.~Adams {\it et al.}  [STAR Collaboration],
  Phys.\ Rev.\ Lett.\  {\bf 92}, 052302 (2004)

\bibitem{Adler:2003kt}
  S.~S.~Adler {\it et al.}  [PHENIX Collaboration],
  Phys.\ Rev.\ Lett.\  {\bf 91}, 182301 (2003)
\bibitem{Muller:2005pv}
  B.~Muller, R.~J.~Fries and S.~A.~Bass,
  Phys.\ Lett.\ B {\bf 618}, 77 (2005)

\bibitem{Yang:2005dg}
  C.~B.~Yang,
  J.\ Phys.\ G {\bf 32}, L11 (2006)

\bibitem{Greco:2004ex}
  V.~Greco and C.~M.~Ko,
  Phys.\ Rev.\ C {\bf 70}, 024901 (2004)
\bibitem{Molnar:2005wf}
  D.~Molnar,
  arXiv:nucl-th/0512001.


\bibitem{Pratt:2004zq}
  S.~Pratt and S.~Pal,
  Nucl.\ Phys.\ A {\bf 749}, 268 (2005)
  [Phys.\ Rev.\ C {\bf 71}, 014905 (2005)]


\bibitem{Molnar:2004rr}
  D.~Molnar,
  arXiv:nucl-th/0408044.


\bibitem{Greco:2005jk}
  V.~Greco and C.~M.~Ko,
  arXiv:nucl-th/0505061.

\bibitem{Lin:2002rw}
  Z.~w.~Lin and C.~M.~Ko,
  Phys.\ Rev.\ Lett.\  {\bf 89}, 202302 (2002)


\bibitem{Scheibl:1998tk}
  R.~Scheibl and U.~W.~Heinz,
  Phys.\ Rev.\ C {\bf 59}, 1585 (1999)

\bibitem{Oldenburg:2005er}
  M.~Oldenburg  [STAR Collaboration],
  arXiv:nucl-ex/0510026.
\bibitem{Dong:2004ve}
  X.~Dong, S.~Esumi, P.~Sorensen, N.~Xu and Z.~Xu,
  Phys.\ Lett.\ B {\bf 597}, 328 (2004)

\bibitem{phenix}
  A. Adare {\it et al.}  [PHENIX Collaboration],
arXiv:nucl-ex/0608033; M.~Issah, talk in AGS User meetings,
  http://www.bnl.gov/rhic\_ags/users\_meeting/Workshops/5.asp,
  http://www.phenix.bnl.gov/phenix/WWW/publish/missah/
  RHIC-AGS-Users06/rhic-ags\_missah.ppt
\bibitem{Kolb:2000sd}
  P.~F.~Kolb, J.~Sollfrank and U.~W.~Heinz,
  Phys.\ Rev.\ C {\bf 62}, 054909 (2000)

\bibitem{Danielewicz:1994nb}
  P.~Danielewicz,
  Phys.\ Rev.\ C {\bf 51}, 716 (1995)

\bibitem{Csanad:2003qa}
  M.~Csanad, T.~Csorgo and B.~Lorstad,
  Nucl.\ Phys.\ A {\bf 742}, 80 (2004)

\bibitem{jjia:derivation}
For a thermal source and at $p_T \ll mv_T$, where $v_T$ is the
radial flow velocity,
\begin{eqnarray}
f(p,x) \propto \exp(-\frac{\gamma_T(m_T-\overrightarrow{p_T} \cdot
\overrightarrow{v_T}) \pm\mu}{T})\\\nonumber \propto
\exp(-\frac{\gamma_Tm_T}{T}) \propto
\exp(-\frac{m\left<v(\phi)\right>^2}{2T})
\end{eqnarray}
we have ignored the $\phi$ dependence of the $T$ and $\mu$. For
this simple scenario, the elliptic flow would be,
\begin{equation}
v_2 \approx
\frac{m\left[\left<v\right>_y^2-\left<v\right>_x^2\right]}{8T}
=\frac{p_T^2\epsilon_v}{8mT}
\end{equation}
$\epsilon_v =
\left(\left<v\right>_y^2-\left<v\right>_x^2\right)/\left<v\right>^2$
is the velocity anisotropy and should to first order be
proportional to spacial anisotropy~\cite{Kolb:2000sd}: $\epsilon_v
= A\epsilon_x +O (p_T)$.


\bibitem{Huovinen:2001cy}
  P.~Huovinen, P.~F.~Kolb, U.~W.~Heinz, P.~V.~Ruuskanen and S.~A.~Voloshin,
  Phys.\ Lett.\ B {\bf 503}, 58 (2001)


\end{references}
\end{document}